\documentclass{IEEEcsmag}

\usepackage[colorlinks,urlcolor=blue,linkcolor=blue,citecolor=blue]{hyperref}

\usepackage{upmath}

\usepackage[para,online,flushleft]{threeparttable}

\jmonth{May/June}
\jname{MultiMedia}
\pubyear{2020}

\begin{document}


\title{Building a Manga Dataset "Manga109" with \\
Annotations for Multimedia Applications}

\author{Kiyoharu Aizawa, Azuma Fujimoto, Atsushi Otsubo, Toru Ogawa, Yusuke Matsui, Koki Tsubota, Hikaru Ikuta}
\affil{The University of Tokyo}

\markboth{Manga109}{Paper title}

\begin{abstract}
Manga, or comics, which are a type of multimodal artwork, have been left behind in the recent trend of deep learning applications because of the lack of a proper dataset. Hence, we built Manga109, a dataset consisting of a variety of 109 Japanese comic books (94 authors and 21,142 pages) and made it publicly available by obtaining author permissions for academic use. We carefully annotated the frames, speech texts, character faces, and character bodies; the total number of annotations exceeds 500k. This dataset provides numerous manga images and annotations, which will be beneficial for use in machine learning algorithms and their evaluation. In addition to academic use, we obtained further permission for a subset of the dataset for industrial use. In this article, we describe the details of the dataset and present a few examples of multimedia processing applications (detection, retrieval, and generation) that apply existing deep learning methods and are made possible by the dataset.
\end{abstract}

\maketitle

\chapterinitial{Manga}, or comics, are popular worldwide. The digital manga market is growing approximately 10\% every year in Japan. Manga is a type of multimedia that has a unique representation style -- a binary hand-drawn image and speech texts comprise a frame and the layout of the frames represents a sequence of scenes. 
From the point of view of multimedia research, multimedia processing of manga has been performed much less often than that of natural images. One of the major reasons is the lack of a proper dataset of 
high quality manga made by professional authors that is freely available to researchers. It is challenging to negotiate copyright with publishers and authors as well as to obtain permission when publishing papers.

\begin{figure}[h!] 
\includegraphics[width=7cm]{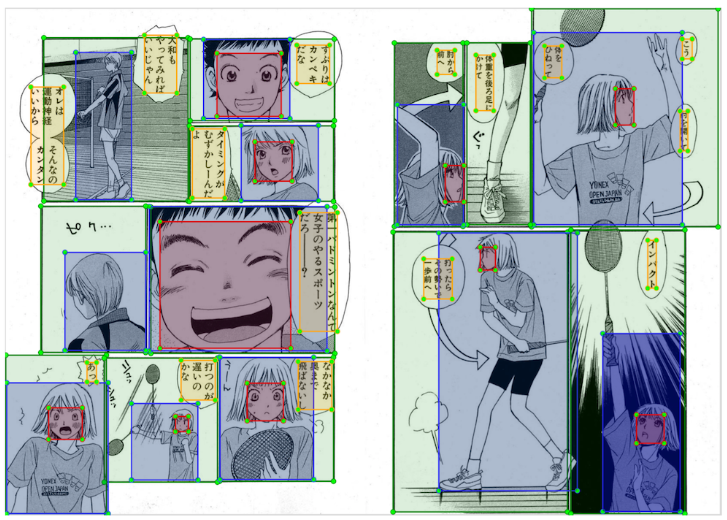}
\caption{Sample page from Manga109. Frames, texts, character faces, and bodies are annotated with their bounding boxes. YamatoNoHane \copyright Saki Kaori}
\label{fig:1}
\end{figure}

\begin{table*}[h!]
  \caption{Comic datasets and Manga109}
  \label{table:dataset}
  \centering
  \begin{threeparttable}
    {\tabcolsep=2mm \begin{tabular}{@{}llrr|rrrr@{}}
      \hline
      &&&&\multicolumn{4}{c}{Annotations} \\ 
      Dataset&&\#volume&\#page&\#frame&\#text&\#face&\#body \\ \hline
      eBDtheque~\cite{eBDtheque2013}&&25&100&850&1,092 \tnote{$a)$} ~&\textemdash&1,550 \rule[0mm]{0mm}{4mm} \\
      COMICS~\cite{IyyerComics2017}&&3,948&198,657&\small{(1,229,664)} \tnote{$b)$}&\small{(2,498,657)} \tnote{$b)$}&\textemdash&\textemdash  \rule[0mm]{0mm}{4mm}  \\ \hline
      Manga109~\cite{matsui2017sketch}&&109&21,142&\textemdash&\textemdash&\textemdash&\textemdash  \rule[0mm]{0mm}{4mm}  \\
      Manga109~(with annotations) & &109&20,260 \tnote{$c)$}~~&103,900&147,918&118,715&157,152 \rule[0mm]{0mm}{4mm}  \\ \hline
    \end{tabular}}
    \begin{tablenotes}[para]
    \item[$a)$] the number of speech balloons. eBDtheque has bounding boxes of text lines as well.\\
    \item[$b)$] Pseudo-annotation (automatically annotated). \\
    \item[$c)$] Equivalent to 10,130 double-sided pages. 
    \end{tablenotes}
  \end{threeparttable}
\end{table*}

To address this issue, we previously created a dataset of manga, namely the Manga109 dataset \cite{manga109, matsui2017sketch}. This dataset consists of a variety of 109 manga titles of 94 authors, and comprises 21,142 pages in total. It was made publicly available for academic research purposes with proper copyright notation. In this study, we further created annotations consisting of four elements, namely, frames, texts, character faces, and character bodies, and the total number of annotations 
exceeds 500k. Figure \ref{fig:1} shows example pages from the Manga109 dataset. 
Numerous manga images and annotations of this dataset are beneficial for use in machine learning 
algorithms and evaluations. A subset (87 volumes) of the dataset is made available for industrial use as well. 
In this article, we do not discuss novel technical methodologies, but we present the details of the Manga109 together with a few applications made possible by the use of the dataset. 
Manga109 dataset in \cite{matsui2017sketch} only contains image data. In this article, we present details of annotations we have created, which are valuable for various applications. 

In the following, we introduce the status of manga and comics datasets used in research in Section 2,
and present the outline of Manga109 in Section 3. We define the annotations and detail the 
work process of annotation in Section 4. We also detail the subset of the Manga109 for industrial use in Section 5.
In Section 6, we show examples of multimedia research using the Manga109 dataset -- text detection, 
sketch-based retrieval, and character face generation, and we conclude the paper in Section 7.

\section{MANGA and COMIC DATASETS}

We review the currently available comics datasets, which are listed in Table \ref{table:dataset}. Because of copyright issues, the number of datasets of comics that can be used for academic research is very limited. 
Guerin et al. published eBDtheque \cite{eBDtheque2013}. 
This is the first comic dataset that has been released publicly. The dataset consists of a total of  100 pages of French, American, and Japanese comics. 
The pages are chosen from various sources and each page contains details of source information as well. 
Since the number of pages is very small, it is difficult to use this dataset for purposes of machine learning.
Although the number of pages is small, it includes not only the position of frames, speech balloons, and text lines but also their attributes such as the style of balloons (normal, cloud, spike,...), 
the direction of balloon's tail, the reading order of panels, and so on.

Iyyer et al. published COMICS \cite{IyyerComics2017}, which is the largest comics dataset for research purposes. This dataset contains 3,948 volumes of American comics published during the
``Golden Age'' of American comics (1938 -- 1954) obtained from the Digital Comics Museum, which hosts user-uploaded scans of many comics by lesser-known Golden Age publishers that are now in the public domain due to copyright expiration. They have provided bounding box annotations for frames and text. However, most of the annotations were automatically created, and would not be suitable for training or evaluation -- only 500 pages are manually annotated, and the annotations attached to the rest pages are the result of the detection of a Faster R-CNN \cite{ren2015faster} trained on the manually annotated 500 pages.

As for Manga109, we published the images in \cite{manga109,matsui2017sketch} and report the annotations in this study --  this database contains 109 volumes (21,142 pages) of Japanese comics (manga) drawn by 94 authors. These manga have two desirable characteristics: (a) high quality -- all volumes were drawn by professional authors and were professionally published -- and (b) diversity -- the years of publication range from the 1970s to the 2010s, and 12 genres (e.g., sports and romantic comedy) are covered. 

A metadata framework for manga was presented in \cite{nagamori2009}, in which a hierarchical structure of the entities of manga is conceptually described. This annotation framework defined three categories of manga objects which are visual objects, dialogs (texts) and symbols. The visual object category consists of characters, items and scene texts. The dialog (text) category has texts of speech, thought, narration and monologue. The symbol category has lines, onomatopeia and marks. It may be an ideal annotation for manga, but the complexity of the annotation definitions is too high for real data in practice. As described later, we defined a simplified annotation framework.

In the following sections, we describe Manga109 and its annotations.

\begin{table}[b]
  \begin{center}
  \caption{Number of comics volumes in the each decades or the genres in Manga 109}
  \label{table:statistics}
	    \begin{tabular}{|l|r|} \hline
              Era & \# volume \rule[0mm]{0mm}{3mm}\\ \hline \hline
              1970s & 7\rule[0mm]{0mm}{3mm} \\  \hline 
              1980s & 24\rule[0mm]{0mm}{3mm}\\ \hline
              1990s & 45\rule[0mm]{0mm}{3mm}\\ \hline
              2000s & 32\rule[0mm]{0mm}{3mm} \\ \hline
              2010s & 1 \rule[0mm]{0mm}{3mm}\\ \hline
            \end{tabular}
\vspace*{5mm}
      
            \begin{tabular}{|l|r||l|r|} \hline
              Genre & num. & Genre & num. \rule[0mm]{0mm}{3mm} \\ \hline \hline
              Animal & 5 & Humor & 15 \rule[0mm]{0mm}{3mm} \\ \hline
             Battle & 9 & Love romance & 13 \rule[0mm]{0mm}{3mm} \\ \hline
              Fantasy & 12 & Romantic comedy & 13 \rule[0mm]{0mm}{3mm} \\ \hline
              4 frame cartoons & 5 & Science fiction & 14 \rule[0mm]{0mm}{3mm}  \\ \hline
              Historical drama & 6 & Sports & 10 \rule[0mm]{0mm}{3mm} \\ \hline
              Horror & 2 & Suspence & 5 \rule[0mm]{0mm}{3mm} \\ \hline
            \end{tabular}
  \end{center}
\end{table}

\section{Manga109}

A large-scale dataset of manga images is crucial for manga research. In early studies on manga image processing, fair comparisons of methods could not be conducted because of the lack of a sufficiently large dataset. Manga is artwork, and copyright is a sensitive issue. If we were to publish our research results using commercial manga that have been professionally published, it would be necessary to obtain the permission of the authors or publishers. In general, this takes time, and permission is not easy to obtain. Thus, to facilitate manga research, it is necessary to build a manga dataset that is publicly available to the academic research community. The availability of this dataset will accelerate multimedia research on manga.

The Manga109 dataset has addressed the sensitive copyright issues. With the help of Mr. Ken Akamatsu, a professional manga author and the founder of J-Comic Terrace Corporation, we obtained permissions from 94 professional authors for the use of 109 volumes in academic research. 
All the manga titles of Manga109 are in the archive ``Manga Library Z'' \cite{mangaz}, which is run by J-Comic Terrace. It has thousands of titles that are currently out of print. We chose 109 titles from the archive that cover a wide range of genres and publication years. Researchers can use them freely with appropriate citation in their research tasks such as retrieval, localization, character recognition, colorization, text detection, and optical character recognition. As shown in Table \ref{table:statistics}, the manga titles were originally published from the 1970s to the 2010s. The Manga109 dataset covers various categories, including humor, battle, romantic comedy, animal, science fiction, sports, historical drama, fantasy, love and romance, suspense, horror, and four-frame cartoons. 

\section{Manga109 Annotations}

\begin{figure}[h]
  \centering
  \includegraphics[width=7cm]{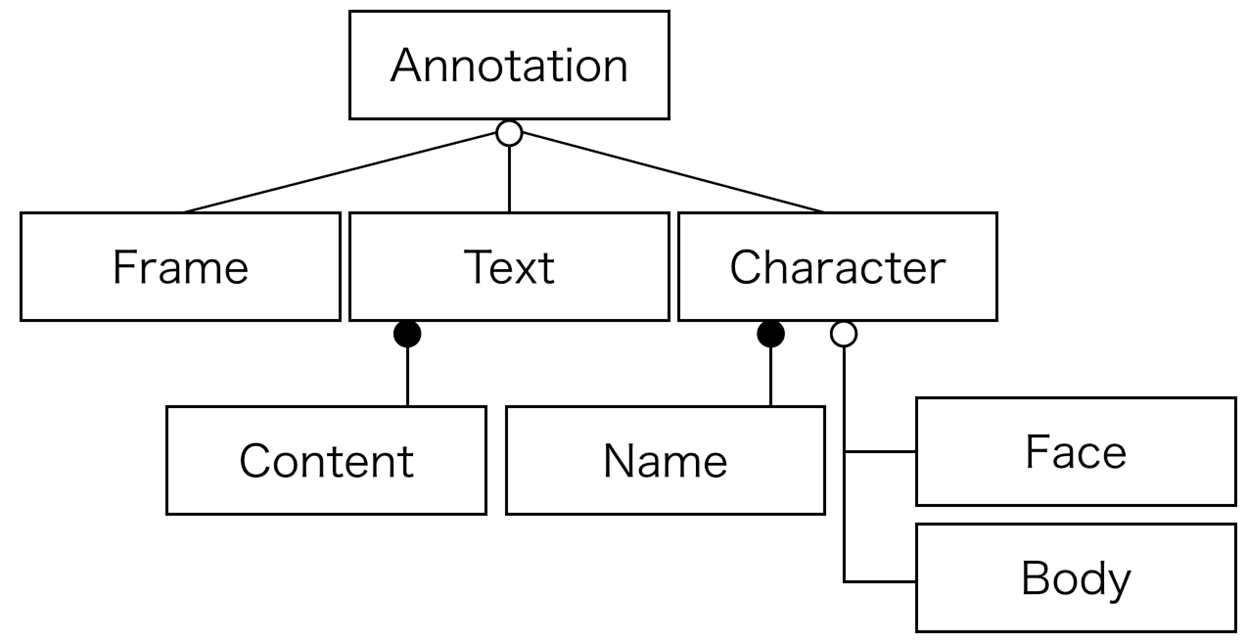}
  \caption{Annotations. We define four different bounding box annotations (Frame, Text, Character  
  Face and  Character Body) and content annotations (Text Content and Character Name). }
  \label{fig:label}
\end{figure}

In this section, we explain Manga109 Annotations, which is a new annotation dataset based on Manga109. Because the original Manga109 does not contain annotations, we defined an annotation framework and manually annotated the entire dataset. 
These annotations contain bounding box annotations for four different kinds of manga objects -- frames, text, character faces, and character bodies -- and content annotations -- character names and text contents -- as shown in Fig. \ref{fig:label}.
With Manga109 Annotations, one can easily
train and evaluate systems for manga object detection, retrieval, character recognition, and other tasks. 

\begin{figure}[!t]
  \centering
  \includegraphics[width=0.7\hsize]{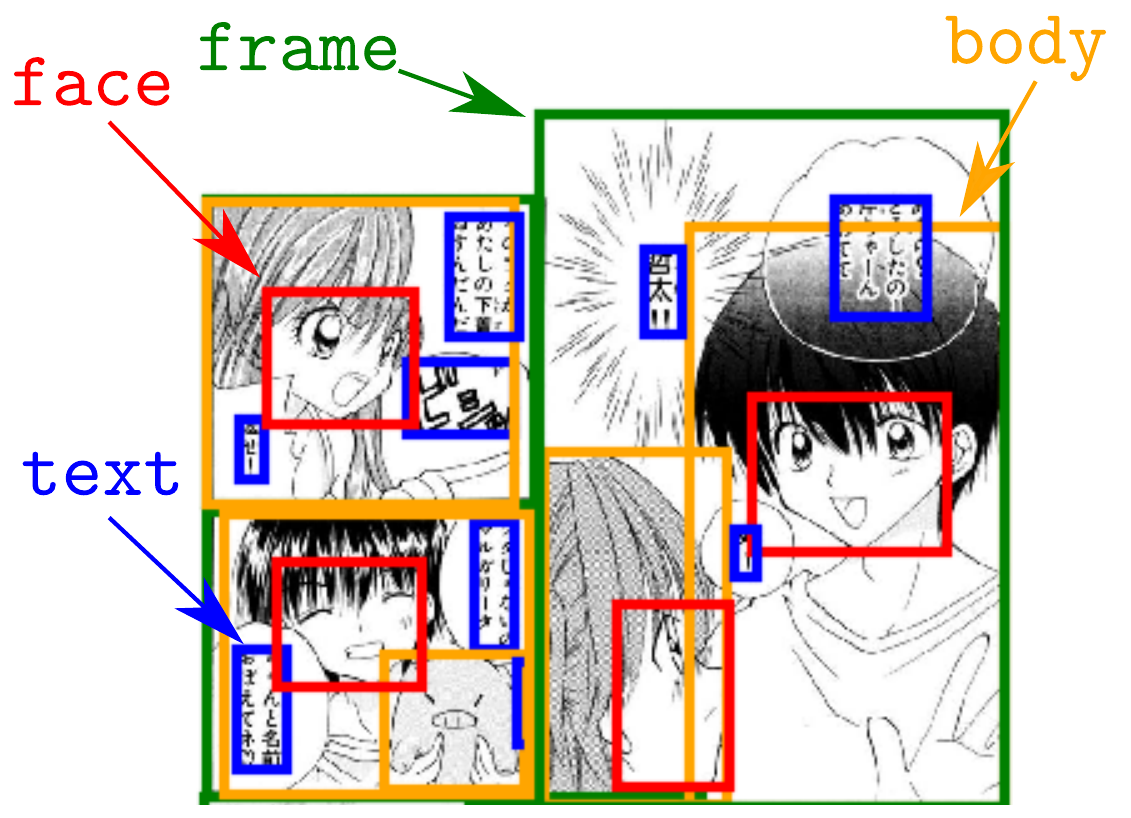}
  \caption{Four object categories.
 These elements are fundamental and play important roles in comics.
  ``BakuretsuKungFuGirl'' \textcopyright Ueda Miki }
  \label{fig:four-annotation}
\end{figure}

\subsection{Bounding box annotations}
The bounding box annotation consists of a bounding box and a category label. The bounding box consists of four values (xmin; ymin; xmax; ymax), which represents a rectangular region in an image. The category label consists of one of the following categories: frame, text, face, or body. We selected these four categories because these elements play important roles in comics, as explained in the paragraphs that follow. 

Figure \ref{fig:four-annotation} illustrates an example of the annotations. A bounding box is indicated by a rectangle, and its color indicates the category label. In the annotation process, human annotators selected each region that belongs to one of these categories with a bounding box and assigned a category label to that region.

\begin{itemize}
\item Frame:\\
A frame is a region of a page that describes a scene.  
Typically, a page in comics is composed of a set of frames.
Visual elements (e.g., characters, background, and texts) are usually drawn inside a frame. Readers follow these frames in a certain order to understand the story. A frame is usually indicated by a rectangle with a solid border, but the shape can be any polygon. Because a frame is a basic unit of image representation in comics, frame detection is important. Although several frame detection methods have been proposed to date \cite{Arai2011,Rigaud2013,Rigaud2015}, 
their performances could not previously be compared because of the lack of publicly available ground-truth annotations. Our Manga109 provides this ground truth.

\item Text:\\
Text is an important element that contains a character's speech, a monologue, or narration. Most of texts are placed in bounded white regions and overlaid on characters or background. We call these regions speech balloons. Some texts are placed directly in the image without speech balloons. Text region detection has been presented in \cite{Rigaud2013,aramaki2016,Rigaud2017,Uchida2019}.

\item Face and body:\\
The faces and bodies of characters are the most important drawn objects. We define the region of a face as a rectangle that includes the eyebrows, cheek, and chin. The region of a body is defined as a rectangle that includes body parts such as the head, hair, arms, and legs. Because the face is also a part of the body, the face is always included in the body region. Some comics use animals, such as dogs and cats, as main characters. In these cases, we treat them as we do for human characters. Because of the frames, faces and bodies are often only partially visible. For example, only the upper body of a boy is shown in Fig. \ref{fig:four-annotation}. 

\end{itemize}

\subsection{Content annotations}
We defined two types of content information: character names (IDs) and the contents of the texts. We annotated character names to all faces and bodies. In the 109 volumes in the dataset, 2,979 unique character names exist. For the text annotations, we represent each content of the text as a Unicode string. The annotators were asked to input the content of each text manually. As a result of this annotation, we produced 2,037,046 characters (5.7 MB) of text data. These text data can be quite useful for a number of applications, including automatic translation, text recognition, speaker detection, and multimodal learning.

\begin{figure}[!t]
  \centering
  \begin{minipage}{0.45\hsize}
  \includegraphics[width=1\hsize]{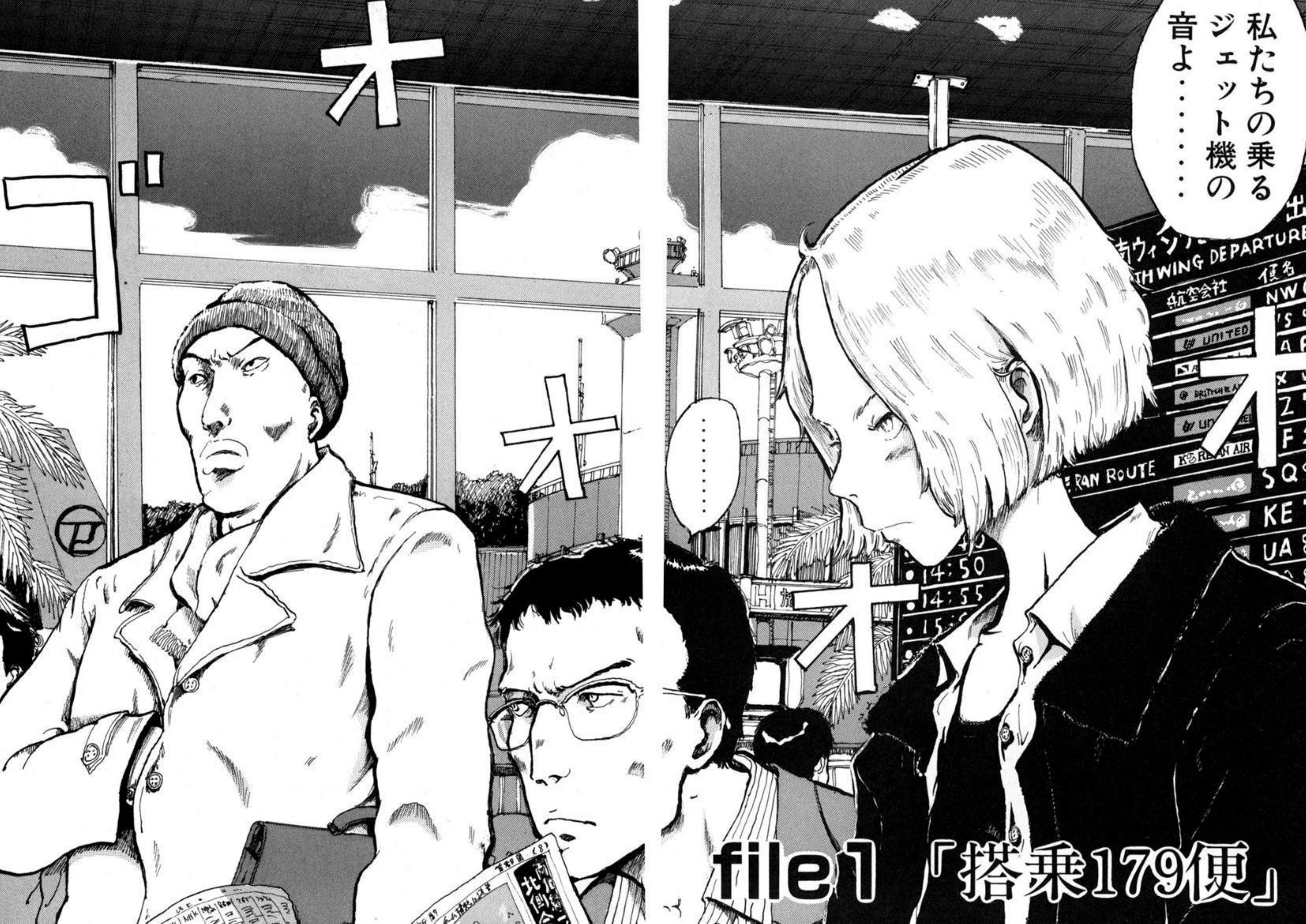}
    \label{fig:dataset:double-sided:a}
  \end{minipage}
  \hfil
  \begin{minipage}{0.45\hsize}
    \includegraphics[width=1\hsize]{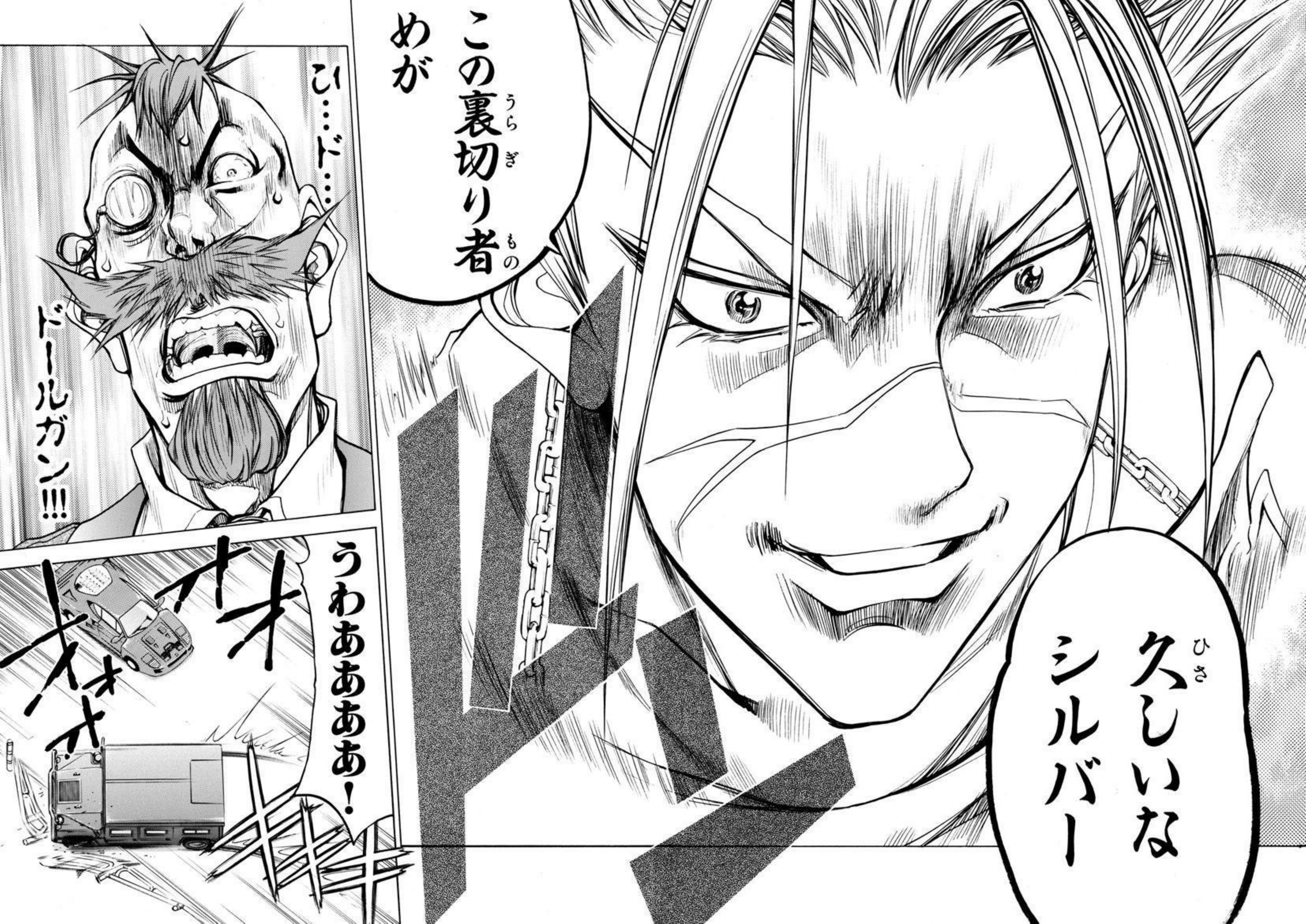}
    \label{fig:dataset:double-sided:b}
  \end{minipage}
  \caption{Double-sided pages.
    In both of these examples, the left and right pages are combined to describe a large image.
    The character in the middle is located across both pages. 
      Left ``HanzaiKousyouninMinegishiEitarou'' \textcopyright Ki Takashi,  
    Right ``DollGun'' \textcopyright Deguchi Ryusei }
  \label{fig:dataset:double-sided}
\end{figure}


\begin{figure*}[t]
  \centering
 \includegraphics[clip, width=14.0cm]{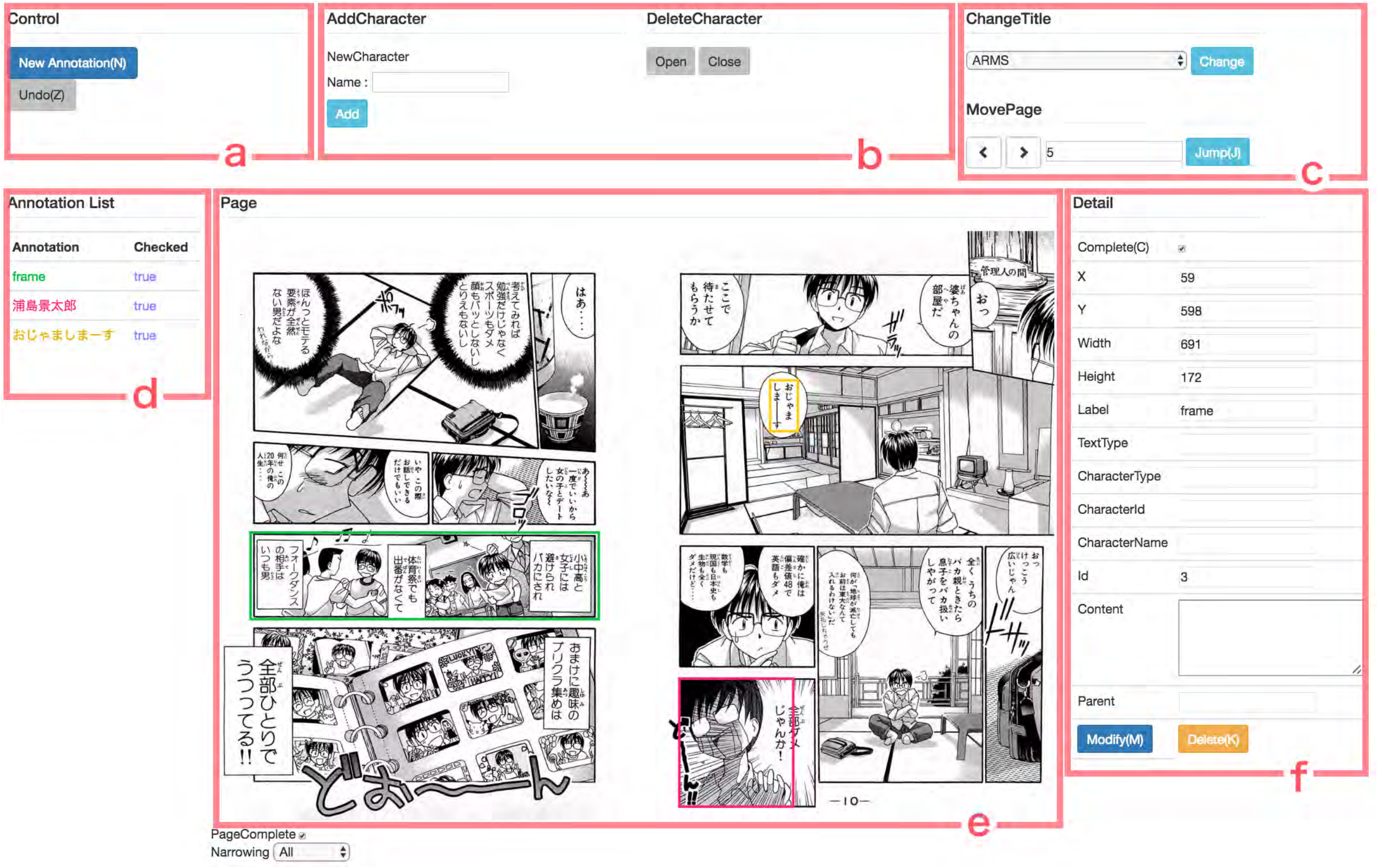}
  \caption{Software window areas: (a) New annotation and undo, (b) Character addition and deletion, (c) Title or page move, (d) Annotation list, (e) Page, (f) Detail}
  \label{fig:window}
\end{figure*}

\subsection{Annotation process}
Here, we describe the procedure of our manual annotation process. Before the annotation, we performed two preprocessing steps over all images. First, we concatenated the left and
right pages into one. We do this because authors sometimes use both left and right pages to 
draw a high-impact picture, as shown in Fig. \ref{fig:dataset:double-sided}.  
We call this pair of pages a ``double-sided page.'' 
This format enables us to annotate objects drawn across both left and right pages. 
To annotate these objects properly, we treated all pages as double-sided pages. Note that this concatenation was conducted without considering the contents of the pages. The size of a concatenated page is typically 1654 $\times$ 1170 pixels. Next, we skipped pages such as cover pages, the tables of contents, and the postscripts. We have done so since they generally do not contain manga content, and the structure of their pages differs from that of regular manga pages. 
After these preprocessing steps, the number of pages to be annotated was 20,260, which is  
10,130 pages in double-page format. 

The annotation was conducted in three steps. First, we invited 72 workers to participate as annotators. Each worker was first assigned one volume of manga. 
If a worker performed well in terms of annotation speed, we assigned additional volumes to that worker. The fastest worker annotated 11 volumes. For this annotation task, we developed a web-based annotation tool. The details of the tool are described in the next section. Because the task is complicated (including steps such as selecting a region, assigning a category label, and typing the contents of texts), we decided to recruit individuals and did not use a crowd-sourcing service, 
which is not reliable enough for this task.
From the beginning, including the design of the annotation framework, this first annotation process took almost 12 months. 

Second, to ensure the quality of the annotations, we double checked all pages manually with the help of 26 workers, all of whom were members of our laboratory. The workers flagged pages with errors that were caused by mistakes and misunderstanding of the annotation criteria. Because comics are complicated documents, there were many errors after the first annotation. As a result, 2,503 double-sided pages were marked as having any errors. 

Finally, we corrected the annotations in the flagged pages. We developed another tool specialized for this correction process. The correction was also conducted by the 26 workers who participated in the double-checking step. These second and third steps consisted of an additional duration of 10 months. 

Table \ref{table:dataset} shows the statistics of the Manga109 annotations dataset. This dataset is the largest manually annotated comics dataset. In particular,  it has face and body annotations. This is a clear advantage because characters are one of the most important targets for the detection task.


\section{Annotation Tool}

We herein summarize the annotation tool we built. The annotation of images requires a considerable amount of effort involving several tens of workers. For this reason, we developed a web-based program, coded using HTML and JavaScript to enable the tool to be easily run on various devices without the need for installation.

We designed the software such that a rectangular area can be easily specified and label assignment can be easily performed. The software window is divided into six areas, as shown in Fig.~\ref{fig:window}. A double-sided page can be viewed in the \textit{page area} (e), and all the annotations for the current page can also be viewed in the \textit{page area} (e) or in the \textit{annotation list} (d).
When the number of annotations increase, the \textit{page area} (e) often becomes confusing due to the increase of the number of bounding boxes. Therefore, the \textit{annotation list} (d) is advantageous in this regard.
New annotations can be added by pressing the "New Annotation" button in the \textit{new annotation and undo area} (a), and 
a rectangular area can be specified in the \textit{page area} (e) and subsequently, a label can be assigned or more information in the \textit{detail area} (f) 
can be provided.
When editing the existing annotations, the target in the \textit{annotation list area} (d) or \textit{the page area} (e) can be selected, 
a bounding box in the \textit{page area} (e) and the information in the \textit{detail area} (f) can be edited.
The operations performed in a page are executed in a single window, without the need for screen transitions or scrolling. 
Keyboard shortcuts can also be used as an alternative to the mouse.

In the correction tool, which was also built by us, the interface is similar to the annotation tool described above. It uses transparent colored rectangles covering the objects for better visual presentation. Unlike the annotation tool, the correction tool is intended for use only on individual or local PCs in the laboratories.

\section{Manga109-s for Industrial Use}

Since we made Manga109 publicly available, we have received many requests to use our dataset from companies. This dataset is unique, and the application of machine learning to the content of manga and comics is commercially valuable. According to the original agreement with authors, the dataset was limited to academic use only and commercial use was not allowed. 
Hence, we re-obtained the permissions of authors so that the dataset can be used by companies as well. Our request to update the agreement was accepted by 75 authors whose comics consist of  87 out of the 109 volumes of the dataset. We named this subset of the dataset  Manga109-s. Companies can use the dataset for their research and development. They can use the results of machine learning for commercial purposes. The details of the conditions for use can be found on the Manga109 website \cite{manga109}.


\section{Examples of Multimedia Processing using the Manga109 Dataset}
 
We briefly present three examples of multimedia processing, which we conducted using the dataset. 
They are text detection, sketch-based retrieval, and character face generation. Because of the large amount of high-quality data, the results obtained by applying existing methods produce substantially satisfactory results. 

There are a wide variety of research conducted on manga and comic processing, 
which appeared in previous MANPU workshops \cite{manpu}. For more examples, 
readers can refer to the works therein.

\begin{figure}[h!] 
\centering
\includegraphics[width=7cm]{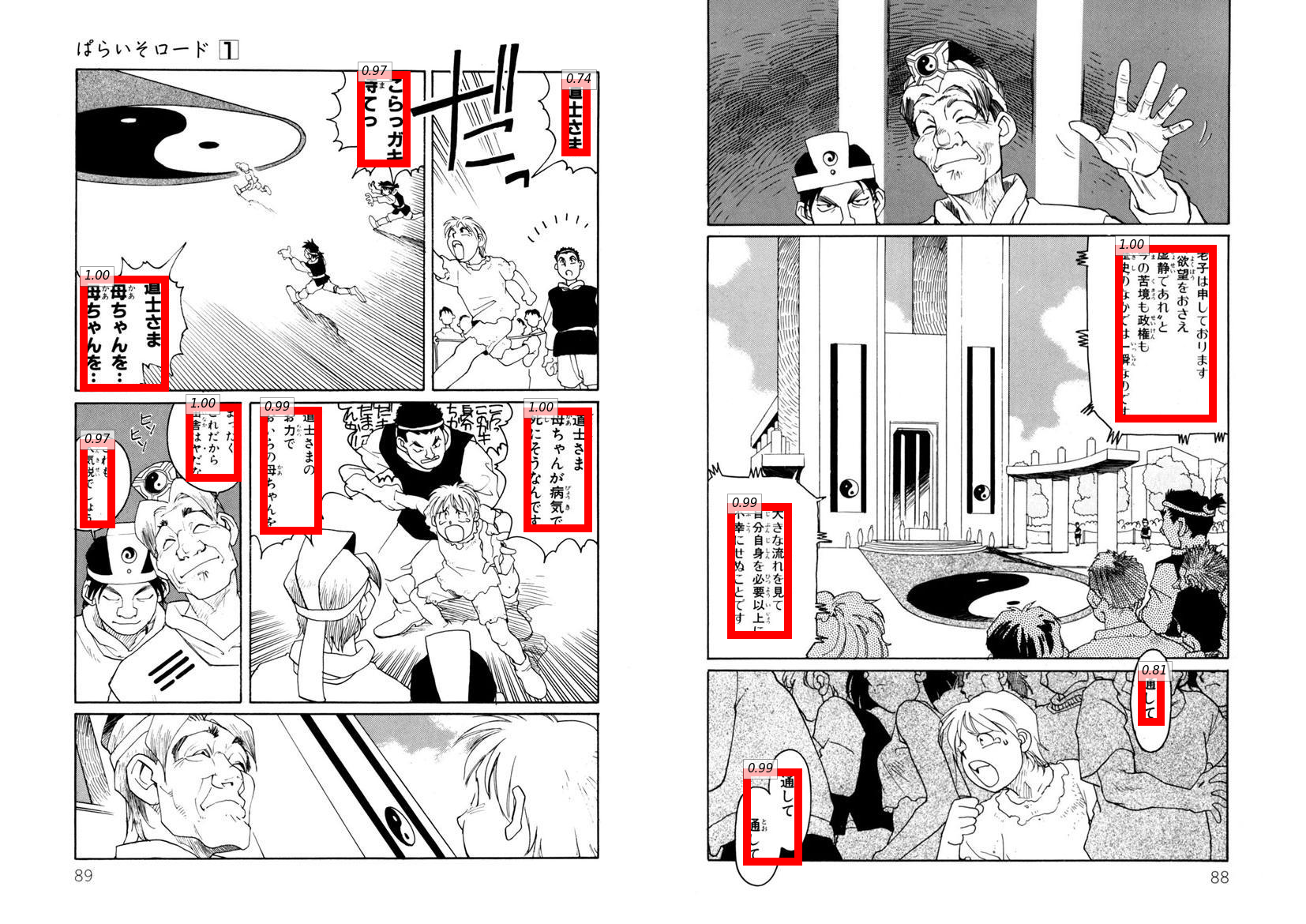} \\
\vspace{5mm}
\includegraphics[width=7cm]{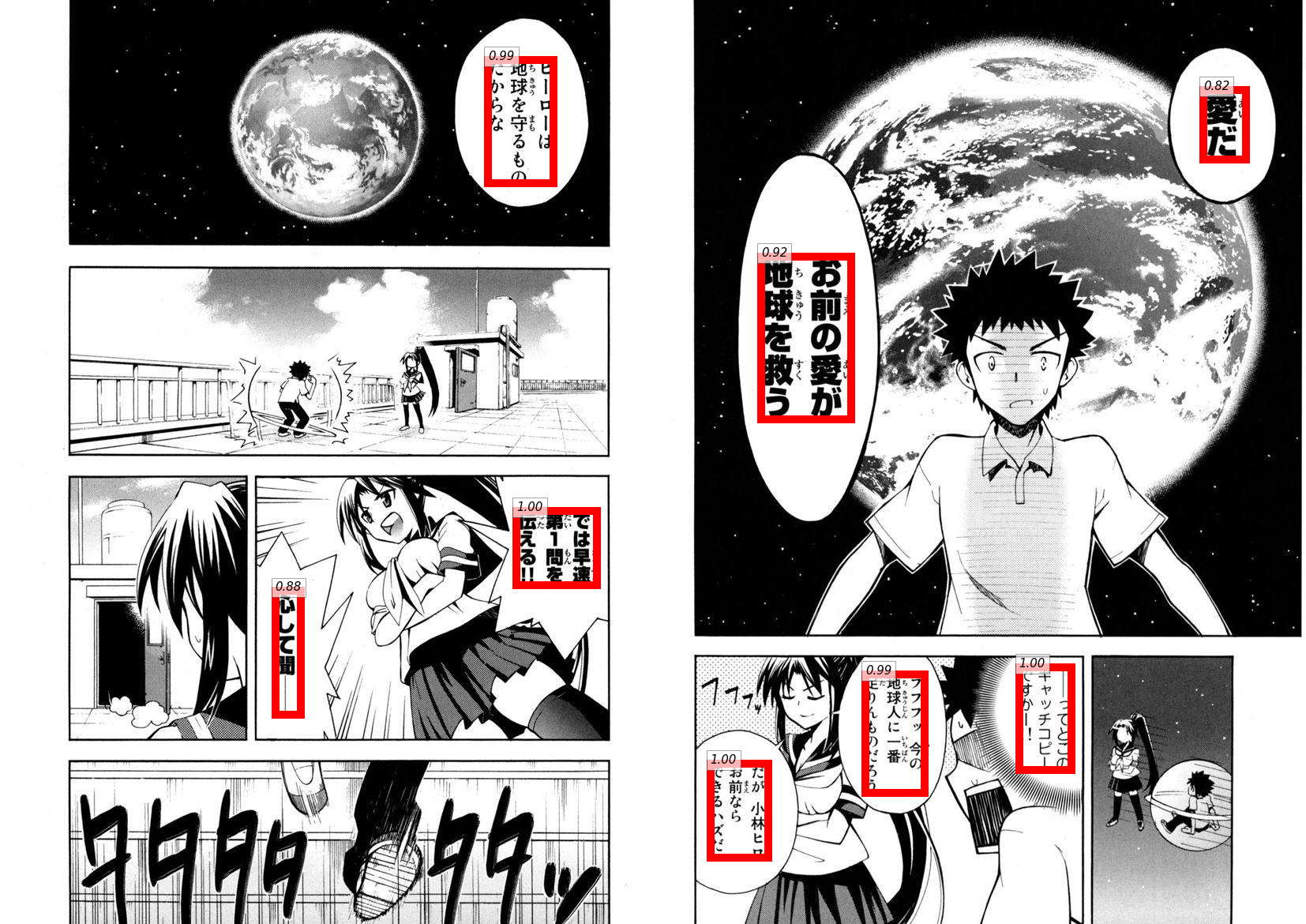}
\caption{Text detection results. The red bounding boxes indicate the detection results. 
The numbers on the bounding boxes show the confidence scores. 
Almost all of them are above 0.9 and the lowest is 0.74.
Top row: ``ParaisoRoad'' \textcopyright Sugano Hiroyuki.
Bottom row: ``MeteoSanStrikeDesu'' \textcopyright Takuji. }
\label{fig:detection}
\end{figure}

\begin{table}[b]
  \begin{center}
  \caption{Text detection accuracies under different sized training data}
  \label{table:comparison}
	    \begin{tabular}{|l|r|l|r|} \hline
Training data size & AP & Training data size & AP \rule[0mm]{0mm}{3mm}\\ 
 \hspace{5mm} (SSD512) &  & \hspace{5mm} (SSD300) & \rule[0mm]{0mm}{3mm}\\ \hline 
      10 volumes & 84.9 &  10 volumes & 80.3 \rule[0mm]{0mm}{3mm}\\ \hline 
      30 volumes & 88.0  & 30 volumes & 84.2 \rule[0mm]{0mm}{3mm}\\ \hline
      99 volumes & 91.8  & 99 volumes & 88.9 \rule[0mm]{0mm}{3mm}\\ \hline
  \end{tabular}
  \end{center}
\end{table}

\subsection{Text detection}
Text detection is one of the most needed processing tasks for manga.
For example, when translating manga into other languages, text detection is the first step.
There have been several methods proposed for the text detection of manga and comics \cite{Rigaud2013, aramaki2016}.
One study used a neural network and evaluated it on the small eBDtheque dataset \cite{Uchida2019}. 
Because those previous studies did not use a large dataset, their results are far from satisfactory.
Using the large amount of data in Manga109, which has more than 147k text bounding boxes, 
we can train standard  object detectors based on neural networks.  

We applied a Single Shot Multibox Detector (SSD)~\cite{liu2016ssd} to detect text in manga images. We trained the SSD for text detection. 
SSD is one of the state-of-the-art object detectors.
We randomly chose 19 of 109 volumes for testing, and the rest 90 volumes for the training. 
The authors and titles of 19 volumes are not included in those of the rest 90 volumes. 
We trained SSD300 and SSD512, which have 300 $\times$ 300 and 512 $\times$ 512 input resolutions, respectively. 
We consider bounding boxes that have an IoU (intersection of union) of over 0.5 with the ground truth to be
successful and other cases to be failure.
The accuracy of object detection is measured by Average Precision (AP). It is a general
metric defined by the area under precision-recall curve. 

We summarize the results of text detection in Table \ref{table:comparison}. The detection results are highly accurate. The AP of the test data obtained by SSD300 is 0.889 and that of SSD512 is 0.918. The larger resolution has better performance since it can detect smaller regions. Examples of the detection results of SSD512 are shown in Fig.\ref{fig:detection}, which demonstrates the high accuracy of the results. In the figure, the numbers indicate the confidence scores of detection. Almost all the texts in these examples are correctly detected with confidence score higher than 0.9, and the lowest confidence scores of the two examples are 0.74 and 0.82, respectively. By setting a proper threshold, they are correctly detected.

This excellent performance is a result of the use of the large high quality dataset.
These APs are far above those reported in \cite{Uchida2019} which used a smaller dataset 
although the datasets are not the same.
In order to investigate the scale effect of the dataset, we made
experiments with training data of 10 and 30 volumes, randomly chosen from the 90 volumes. They are 
1/9 and 1/3 of the training data used in the above experiments.
The results are shown in Table \ref{table:comparison}. It clearly shows that increasing the size of the dataset significantly improves the performance. 

\begin{figure}[h!] 
\centering
\includegraphics[width=7cm]{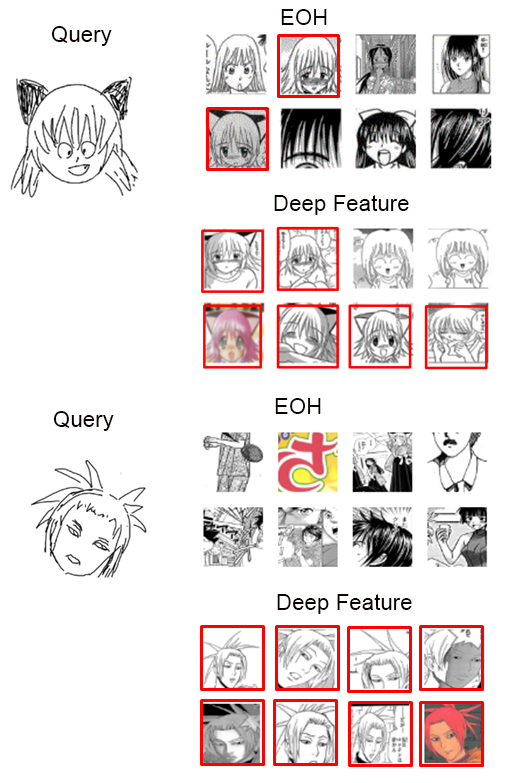} \\
\caption{Comparison of EOH and deep features 
for sketch-based retrieval. The red bounding boxes indicate the correct target character. }
\label{sketch-results} 
\end{figure}

\subsection{Sketch-based manga retrieval} 
Sketch-based manga retrieval was proposed in \cite{matsui2017sketch}. It makes use of the visual features of edge orientation histograms (EOHs), and the retrieval results are very noisy. 

Narita et al. examined a sketch-based manga retrieval method using deep features \cite{narita2017}. They fine-tuned a pre-trained AlexNet for the classification of character face images of Manga109 dataset and used the fc6 layer as a deep feature for retrieval. In the training, they used 200 characters of 70 volumes and more than 70 images per character on average. They removed the screen-tone textures from the manga to simulate sketch images. 

In the retrieval experiment, they used seven volumes of Manga109 and first applied selective search \cite{Uijlings2013} to produce 200 candidates for each double-sided page -- a total of 137,800 candidates over the seven  volumes.
In Fig. \ref{sketch-results}, the retrieval results are compared with those of the previous scheme using EOHs  \cite{matsui2017sketch} and the method using deep features. The results show the top-8 results of the retrieval, and the characters with the red boxes are the ones that match the sketches drawn by the users. As shown in this example, the retrieval results using deep features are substantially better than those of the EOHs. 
In their experiments, they used three different handwriting sketches for these two characters in the figure. 
AP is computed for each sketch, and mAP is the average of them for each character. mAP is reported as 0,26, 0.18 by deep features, and 0.09 and 0.00 by EOH based features.  Use of deep features has quantitatively significantly 
improved performance compared to the EOH based method. 
Even the incorrect results obtained by the deep features are partially correct, in the sense that they are all similar face images. By contrast, the incorrect results obtained using EOHs are completely different from the query face.


\subsection{Character face generation using a GAN}
Generative adversarial networks (GANs) have become a popular framework for generating images. 
Using a vast number of training images,
the latest GAN-based techniques can generate high-quality results.  

\begin{figure}[h!] 
\centering
\includegraphics[width=7cm]{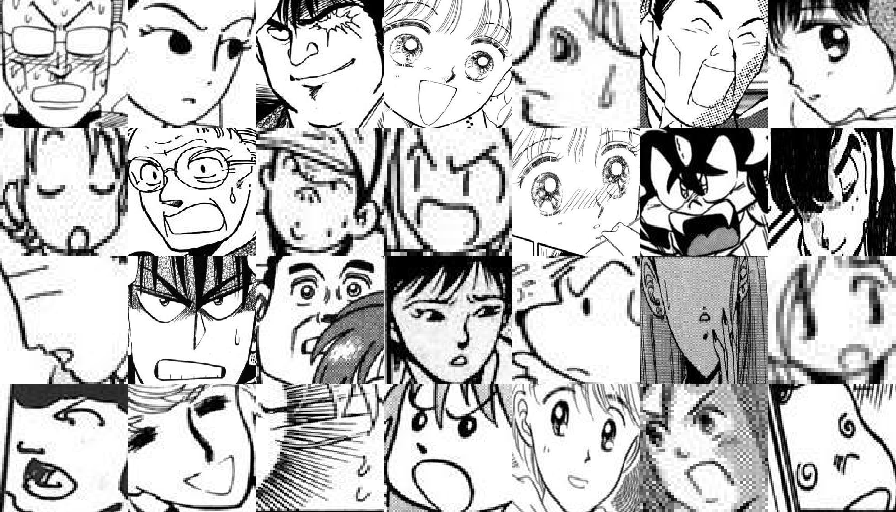} \\
(a) \\
\vspace{5mm}
\includegraphics[width=7cm]{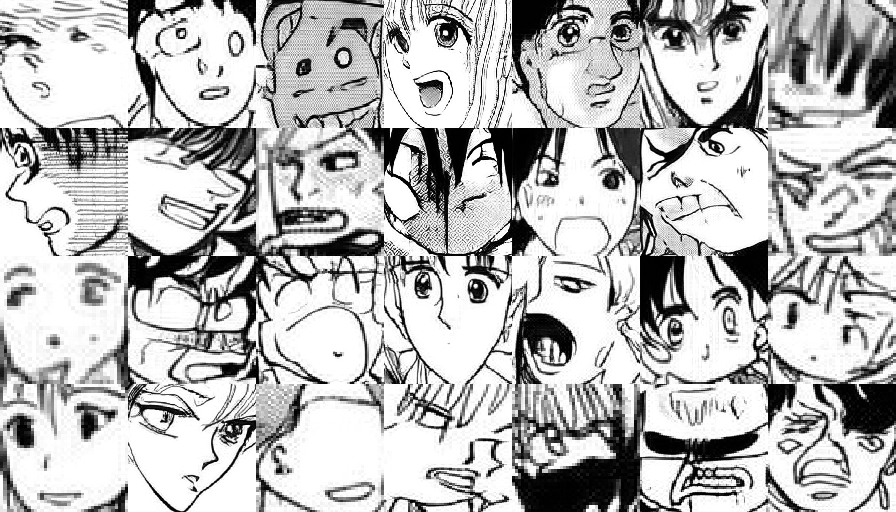}\\
(b) 
\caption{
Results of character face generation by using PGGAN. (a) shows faces used for
training and (b) shows those obtained by generation.}
\label{fig:generation}
\end{figure}

We generated manga character faces using progressive growing of GANs (PGGAN) \cite{PGGAN2018}. The number of character faces in the Manga109 dataset is 118,715. Among them, we chose 104,977 face images with a shorter dimension of more than 30 pixels. In the experiments, we resized all the face images to 128 $\times$ 128 pixels, and trained the network with approximately 114 epochs (a total of 12 million images). We show examples of the training face images and the generated face images in Figs. \ref{fig:generation}(a) and (b). Although each face image seems like it would be difficult to generate -- the character faces are abstract line drawings and we did not apply any alignment preprocessing -- we can generate very reasonable faces. 
Such automatically generated faces can be utilized as background characters or as a guide for creation of new characters, which can support author's artwork.

\section{Conclusions}
 Well-crafted image datasets have played critical roles in evolving image processing technologies, e.g., the PASCAL VOC datasets,  
which enabled image recognition in the 00s, and ImageNet, 
which enabled recent rapid progress in deep architecture. 
In this paper,  we presented Manga109 Annotations, a dataset of a variety of 109 Japanese comic books
 (94 Authors, 21,142 pages) for the research community to use without the need to obtain the permission of authors. We also explained the annotation framework and annotation process to
build Manga109 Annotations. For commercial use, we further obtained the permission of authors for 87 volumes.
We also briefly introduced multimedia processing examples -- text detection, sketch based retrieval, and manga character face generation -- which are easily made possible using our dataset.
We hope the Manga109 Annotations dataset will contribute to the further development of manga research.

\section*{Acknowledgements}
We thank Mr. Ken Akamatsu and J. Comic Terrace for their great help in
building the Manga109 dataset. We thank the authors who kindly gave us their permission
to use their work in the dataset.  We also thank the workers who were engaged in
the annotation process. This work was partially supported by the Strategic Information and
Communications R \& D Promotion Programme (SCOPE) and JSPS 17K19963.

\end{document}